%% file: Path2N3LO.tex
\documentclass[12pt]{article}

\usepackage{hyperref}
\usepackage{cite}
\usepackage{units}
\usepackage{epsfig}
\usepackage{pifont}
\usepackage{amsmath,amsfonts,amssymb,amsthm,mathrsfs}
\usepackage{graphicx,chngcntr,slashed}
\usepackage{cancel}
\usepackage{pstricks}
\usepackage{afterpage}
\usepackage{braket}
\usepackage{caption}
\usepackage{subcaption}
\usepackage{xcolor}
\usepackage{verbatim}
\usepackage{multirow}
\usepackage{tabularx,caption}
\usepackage{float}
\usepackage{afterpage}
\usepackage{comment}
\makeatletter
\def\section{\@startsection {section}{1}{\z@}{+3.0ex plus +1ex minus
  +.2ex}{2.3ex plus .2ex}{\large\bf\boldmath}}
\def\subsection{\@startsection{subsection}{2}{\z@}{+2.5ex plus +1ex
minus +.2ex}{1.5ex plus .2ex}{\normalsize\bf\boldmath}}
\def\subsubsection{\@startsection{subsubsection}{3}{\z@}{+3.25ex plus
 +1ex minus +.2ex}{1.5ex plus .2ex}{\normalsize\it}}

\def\beq{\begin{equation}}
\def\eeq{\end{equation}}
\newcommand{\refcite}[1]{ref.~\cite{#1}}

\newcommand{\eq}[1]{eq.~\eqref{eq:#1}}
\newcommand{\eqs}[2]{eqs.~\eqref{eq:#1} and \eqref{eq:#2}}
\newcommand{\nn}{\nonumber}

\newcommand{\cO}{\mathcal{O}}
\newcommand{\cI}{\mathcal{I}}
\newcommand{\df}{\mathrm{d}}
\newcommand{\cut}{{\rm cut}}
\newcommand{\taucut}{{\tau_\cut}}
\newcommand{\sing}{{\rm sing}}
\newcommand{\Tau}{\mathcal{T}}
\newcommand{\qt}{{\vec q}_T}
\newcommand{\bt}{{\vec b}_T}
\newcommand{\img}{\mathrm{i}}
\newcommand{\lqcd}{\Lambda_\mathrm{QCD}}
\graphicspath{{.}{plots/}}

\begin{document}

\thispagestyle{empty}

\def\thefootnote{\fnsymbol{footnote}}

\begin{flushleft}
 SLAC-PUB-17658, BONN-TH-2022-06, MITP-22-021
\end{flushleft}

\null \vspace{-1.0cm}
\begin{flushright}
  Snowmass'2021 Whitepaper 
\\ \quad \\
\end{flushright}

\vspace{1cm}

\begin{center}

{\Large {\bf The Path forward to N$^3$LO}}
\\[3.5em]
{\large
Fabrizio~Caola$^1$, Wen Chen$^{2}$, Claude~Duhr$^{3}$, Xiaohui~Liu$^{4}$, Bernhard~Mistlberger$^{5}$, Frank~Petriello$^{6}$, Gherardo Vita$^{5}$, Stefan~Weinzierl$^{7}$
}

\vspace*{1cm}
{\sl
$^1$ Rudolf Peierls Centre for Theoretical Physics, University of Oxford,
 Oxford OX1 3PU and 
 Wadham College, Oxford OX1 3PN, UK\\[1ex]
 $^2$ Department of Physics, Zhejiang University, Hangzhou, Zhejiang 310027, China\\[1ex]
$^3$ Bethe Center for Theoretical Physics, Universit\"at Bonn, D-53115, Germany\\[1ex]
$^4$ Center of Advanced Quantum Studies, Department of Physics,
Beijing Normal University, Beijing 100875, China and 
Center for High Energy Physics, Peking University, Beijing 100871, China \\[1ex]
$^5$ SLAC National Accelerator Laboratory, Stanford University, Stanford, CA 94039, USA \\[1ex]
$^6$ Department of Physics \& Astronomy, Northwestern University, Evanston, IL 60208, USA and 
High Energy Physics Division, Argonne National Laboratory, Argonne, IL 60439, USA \\[1ex]
$^7$ PRISMA Cluster of Excellence, Institut f{\"u}r Physik, Johannes Gutenberg-Universit{\"a}t Mainz, D - 55099 Mainz, Germany 
}

\end{center}

\begin{abstract}
    The LHC experiments will achieve percent level precision measurements of processes key to some of the most pressing questions of contemporary particle physics: 
    What is the nature of the Higgs boson? 
    Can we successfully describe the interaction of fundamental particles at high energies? 
    Is there physics beyond the Standard Model at the LHC? 
    The capability to predict and describe such observables at next-to-next-to-next-to-leading order (N$^3$LO) in QCD perturbation theory is paramount to fully exploit these experimental measurements. 
    We describe the current status of N$^3$LO predictions and highlight their importance in the upcoming precision phase of the LHC.
    Furthermore, we identify key conceptual and mathematical developments necessary to see wide-spread N$^3$LO phenomenology come to fruition.
\end{abstract}
\newpage
\setcounter{page}{1}
\setcounter{footnote}{0}

\tableofcontents

%%%%%%%%%%%%%%%%%%%%%%%%%%%%%%%%%%%%%%%%%%%%%%%%%%%%%%%%%%%%%%

% \section{Instructions}

% The white papers are meant to address highlights of where the current field stands in a particular research area and possibly discuss bottle necks that need to be addressed to go beyond where we are. Please note that our working group report is limited to be 10-15 pages max, so we can not cover all the exciting developments in the various research areas that fall under our working group. Our hope is that we can refer to these white papers for the detailed technical advances. 

% The following are possible topics to cover in this white paper which are well aligned with the expertise of the authors. Please feel free to extend them, and feel free to add your collaborators/students/postdocs to the list of authors. Note that precision PDFs is covered in a separate white paper. An extended list of topics covered by our group can be found in {\tt https://snowmass21.org/theory/precision}
% \newpage

\input{Chapters/Introduction.tex}
\input{Chapters/Status.tex}

\input{Chapters/Amplitudes.tex}

\input{Chapters/Subtraction.tex}
\input{Chapters/Challenges.tex}

\input{Chapters/Conclusions.tex}

%%%%%%%%%%%%%%%%%%%%%%%%%%%%%%%%%%%%%%%%%%%%%%%
\section{Acknowledgements}  \label{sec:acc}
FC is supported by the ERC Starting Grant 804394 hipQCD and by the UK Science and Technology Facilities Council (STFC) under grant ST/T000864/1.
BM and GV are supported by the Department of Energy, Contract DE-AC02-76SF00515. XL is supported by the National Natural Science Foundation of China under Grant No.~12175016. F.~P. is supported by the DOE grants DE-FG02-91ER40684 and DE-AC02-06CH11357.

\bibliographystyle{jhep}
\bibliography{refs}

\end{document}

%% file: Chapters/Introduction.tex
\section{Introduction}
\label{sec:intro}
In this article, we outline the status and progress of theoretical developments making percent level precision phenomenology at the LHC a reality. 
The particular advancement discussed here is the ability to describe scattering processes of high interest at the LHC to next-to-next-to-next-to-leading order (N$^3$LO) in QCD perturbation theory.

Percent level phenomenology with LHC measurements is highly motivated.
First and foremost, this level of precision represents the frontier of what can be realized experimentally. 
Universal limitations to precision are imposed experimentally by our ability to determine how many particle collisions occur at the LHC, i.e. by the determination of the interaction luminosity.
Recent studies by ATLAS~\cite{ATLAS:2019pzw} and CMS~\cite{CMS:2021xjt} confirm remarkably that levels of precision of about one percent are achievable.
Similarly, another dominant leading uncertainty, the experimental resolution on the observed energy of particles and hadronic jets in the detectors, is at comparable levels~\cite{CMS:2016lmd,ATLAS:2017bje}.
Statistical limitations of measurement will be overcome for a plethora of observables by the upcoming High Luminosity phase of the LHC (HL-LHC), during which roughly twenty times as much data will be recorded compared to the accumulated amount to date~\cite{Dainese:2019rgk}.

Percent level LHC phenomenology represents the opportunity to study some of the most pressing questions of high energy physics. 
The LHC is our first and currently only window to glimpse directly at interactions of all fundamental particles in the Standard Model occurring at the electroweak energy scale of several hundred GeV.
As such, current and future measurements will allow us to pin down the properties and the nature of the Higgs boson in great detail~\cite{Dainese:2019rgk}. 
Precision LHC measurements allow us to determine the fundamental masses and coupling strengths of electroweak particles. 
Specifically, the LHC will allow us to study a large variety of observables involving Higgs bosons, electroweak bosons, top quarks and hadronic jets to astounding precision.
As a consequence, we can directly probe our understanding of fundamental interactions at high energies and explore the nature of electroweak symmetry breaking.
In particular, in the absence of the discovery of new particles, the precision exploration of LHC physics is of paramount importance.
Small deviations of measurements from our Standard Model (SM) expectation would lead to the most profound implications on our understanding of nature.
As a result, precision phenomenology will play a crucial role in the search for physics beyond the Standard Model.

The successful execution of a percent level phenomenology program relies on our ability to interpret and predict the outcome LHC measurements.
The theoretical foundation for predictions of high energy particle collision experiments is perturbative Quantum Field Theory (QFT). 
However, achieving percent level precision is a formidable task and requires substantive developments and a concerted effort of the theoretical physics community.
One particular key aspect that will be necessary is technology and understanding to perform computations of LHC observables at N$^3$LO in QCD perturbation theory.
In this article, we focus on the current status of N$^3$LO computations and discuss what future developments are necessary in order to have predictions readily available and easy to use for key LHC observables.

At this point N$^3$LO QCD predictions are available for a selected number of inclusive LHC cross sections for the production of colorless final state particles~\cite{Anastasiou:2015ema,Anastasiou:2016cez,Mistlberger:2018etf,Dreyer:2016oyx,Duhr:2019kwi,Duhr:2020kzd,Chen:2019lzz,Dreyer:2018qbw,Duhr:2020sdp,Duhr:2020seh,Duhr:2021vwj}.
Furthermore, some early results for more differential predictions became available~\cite{Banfi2016,Dreyer:2016oyx,Dulat:2017prg,Dulat:2018bfe,Chen:2021vtu,Billis:2021ecs,Chen:2021isd,Camarda:2021ict,Chen:2022cgv}.
While these results cover only a limited set of processes, often in an idealized description, a general picture can be already deduced: 
1) N$^3$LO corrections are typically at the order of several percent.
2) Residual uncertainties due to the truncation of the perturbative expansion after the inclusion of these corrections are at the percent level.
3) Overall, the inclusion of N$^3$LO corrections leads to a significanlty improved description of collider physics observables.
Consequently, the inclusion of N$^3$LO corrections becomes mandatory for a successful percent level precision phenomenology program at the LHC.
We present a brief overview of some current N$^3$LO computations in sec.~\ref{sec:status}.

Achieving N$^3$LO  precision for a large range of LHC observables is an ambitious goal with a multitude of challenges ahead.
Such computations rely on the availability of multi-loop scattering amplitudes.
Associated with the computation of such amplitudes are a multitude of technical and conceptual challenges. 
First, the sheer increase of complexity of computations at higher and higher loop orders stretches our analytic and numerical capabilities, continues to challenge us to develop new techniques and improved understanding of QFT.
Second, the mathematical functions that amplitudes are comprised of are often beyond established domains of mathematics. 
Such functions include multiple polylogarithms, elliptic polylogarithms and generalizations beyond that, which are an active field of theoretical research. 
Efficient numerical and analytic techniques for the computation of scattering amplitudes will be a key ingredient to widespread N$^3$LO phenomenology. 
We discuss the computation of scattering amplitudes and its mathematical constituents in sec.~\ref{sec:amplitudes}.

The combination of scattering amplitudes to cross sections is another highly non-trivial step in the computation of N$^3$LO corrections.
Infrared and collinear singularities in the integration over final state particle momenta yield this process highly non-trivial conceptually and practically.
While so far mainly analytic techniques were used for the computation of N$^3$LO inclusive cross sections, extension to more realistic observables will require a shift in the approach.
Techniques for the computation of fully differential cross sections were established at NNLO in QCD perturbation theory, but their extension to one order higher is highly non-trivial and will be part of substantial research. 
Such techniques include so-called slicing and subtraction algorithms that are each associated with a range of particular features and advantages.
We present an overview of such techniques and discuss advancements needed in the future in sec.~\ref{sec:subtraction}.

Realizing N$^3$LO phenomenology on a larger scale is associated with many further challenges.
First, The backbone of perturbative computations are parton distribution functions (PDFs). 
Such PDFs are currently available at NNLO and increasing their perturbative order by one will require a significant community effort. 
Furthermore, N$^3$LO corrections should be made readily available to wider theoretical and experimental community. 
Creating such easy-to-use computer software is key to the successful application of these computations in collider phenomenology, but it is also a very challenging task.
Moreover, the perturbative description of hadronic observables fails in so-called infrared sensitive regions of phase space and perturbative computations need to be supplemented with all-order resummation.
The development of such resummation techniques in unison with N$^3$LO predictions is vital.
In addition, the combination and integration of high order perturbative corrections into parton shower frameworks is highly desirable to achieve not only improved predictions but also a high degree of user friendliness and applicability.
Another key conceptual development will be the estimation of theoretical uncertainties of QFT predictions. Such estimates present an increased conceptual and statistical challenge as we near the regime where theory uncertainties pose the leading uncertainty on observables and need to be part of research.
We outline briefly a non-exclusive set of such challenges for the precision theory program in sec~\ref{sec:challenges}.

%% file: Chapters/Status.tex
\section{Example phenomenological results for $2\to 1$ N$^3$LO processes and what they taught us}
\label{sec:status}

The first hadron-collider production process computed at N$^3$LO accuracy was the inclusive cross section for the production of a Higgs boson in gluon-fusion (the dominant production channel at the LHC) in the limit where the top-quark is infinitely heavy and can be integrated out~\cite{Anastasiou:2015ema,Anastasiou:2016cez,Mistlberger:2018etf}. Currently, also the inclusive bottom-quark fusion and vector boson fusion (VBF) channels are known at this order (though in the case of VBF, the result is only known in the factorizable approach, where color exchange in the $t$-channel is neglected)~\cite{Duhr:2019kwi,Duhr:2020kzd,Dreyer:2016oyx}. N$^3$LO results are also available for di-Higgs production in gluon-fusion and VBF~\cite{Dreyer:2018qbw,Chen:2019lzz}. Very recently, also inclusive results for neutral- and charged-current Drell Yan production have been completed~\cite{Duhr:2020seh,Duhr:2020sdp,Duhr:2021vwj}. 
So far, mainly fully-inclusive cross section have been computed at N$^3$LO, and only very few differential distributions are available, e.g., the invariant-mass distribution of the color-singlet, or in some cases also its rapidity or transverse momentum distribution~\cite{Dulat:2018bfe,Dulat:2017prg,Cieri:2018oms,Chen:2021vtu,Chen:2022cgv}. While for precision collider phenomenology also differential observables including fiducial phase space cuts are desirable, fully-inclusive cross sections are an important theoretical and phenomenological tool to understand the structure of the higher-order QCD perturbation theory, and they allow us to assess the quality of our current predictions. 
In the following we will perform a comparison and a study of the systematics of N$^3$LO corrections to hadron collider observables. 
This will allow us to assess the phenomenological relevance of  these corrections, and to devise a roadmap for theory computations needed to match the experimental and theoretical predictions at current and future collider experiments. 
We will focus on fully-inclusive cross section predictions for $2\to1$ processes at the LHC with a center-of-mass energy $\sqrt{S}=13$ TeV, because these processes have the same topology and essentially only differ by the quantum numbers of the partons in the initial state. 
As such, they form an ideal set of observables to study the systematics of higher-order corrections at this order in perturbation theory.

\subsection{Systematics of inclusive $2\to1$ processes at hadron colliders}
Since the only known way to obtain reliable predictions for collider observables is perturbation theory,
an important phenomenological question is how well perturbative computations approximate the exact result, or equivalently, what is the size of the missing higher orders. This question is closely connected to the residual dependence on the renormalization scale $\mu_R$ and factorization scale $\mu_F$ introduced by the truncation of the perturbative series. This dependence is unphysical, and it is canceled by the $\mu_F$ and $\mu_R$ dependence of the missing higher orders. The variation of a fixed-order computation with these scales is therefore often taken as a measure of the size of the missing higher orders in perturbation theory.

There is no consensus on how to determine the value these unphysical scales, though they must be chosen so that they correspond to a hard scale of the process, to avoid the appearance of large logarithms that would spoil the convergence of the perturbative series. For a $2\to1$ process the only relevant hard scale is the invariant mass $Q$ of the produced color singlet state, so it is natural pick $\mu_F$ and $\mu_R$ proportional to $Q$. The impact of the missing higher orders is then conventionally estimated by varying $\mu_F$ and $\mu_R$ by a factor of 2 around the chosen central scales.

\begin{figure*}[!b]
\centering
\includegraphics[width=0.45\textwidth]{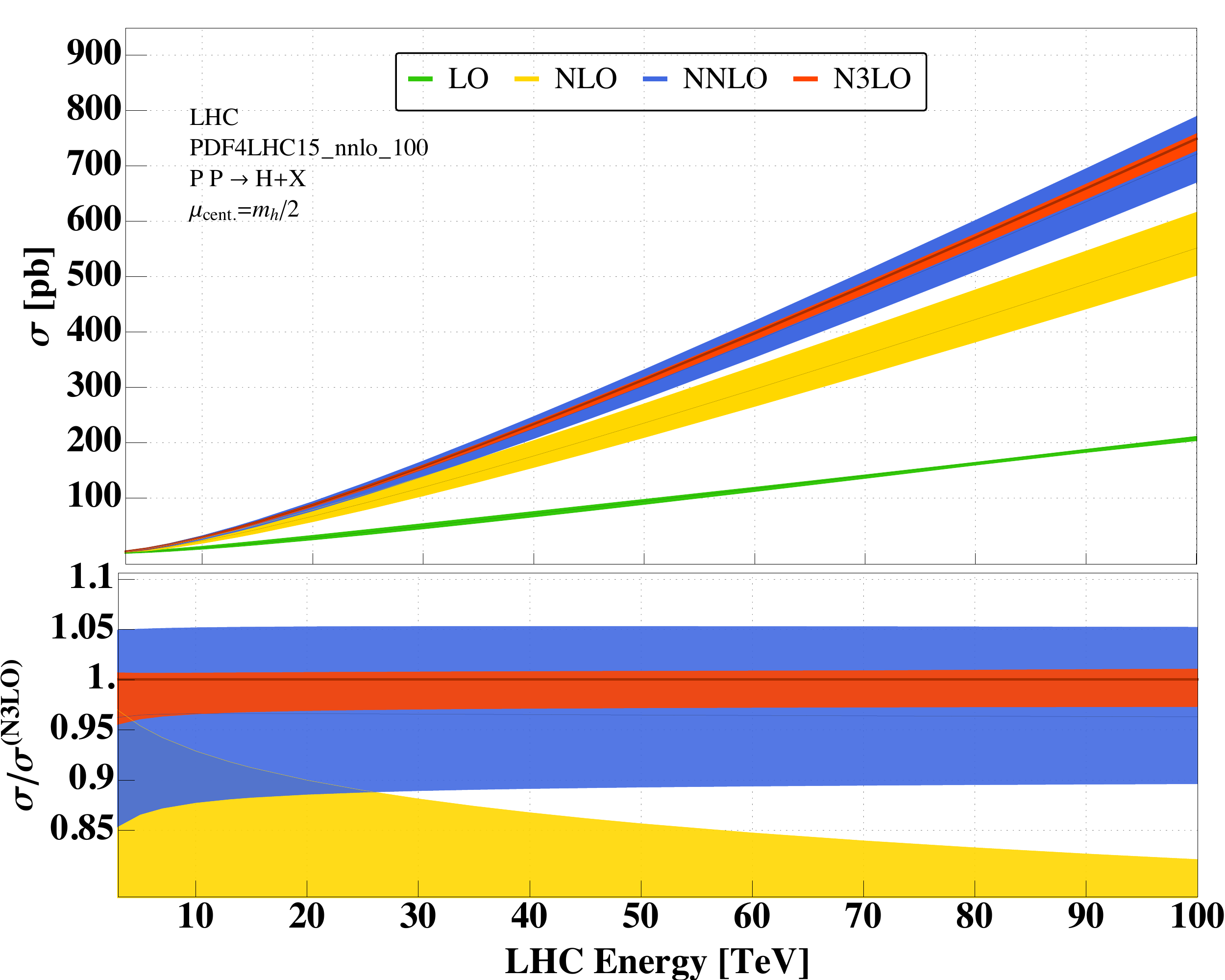}
\includegraphics[width=0.45\textwidth]{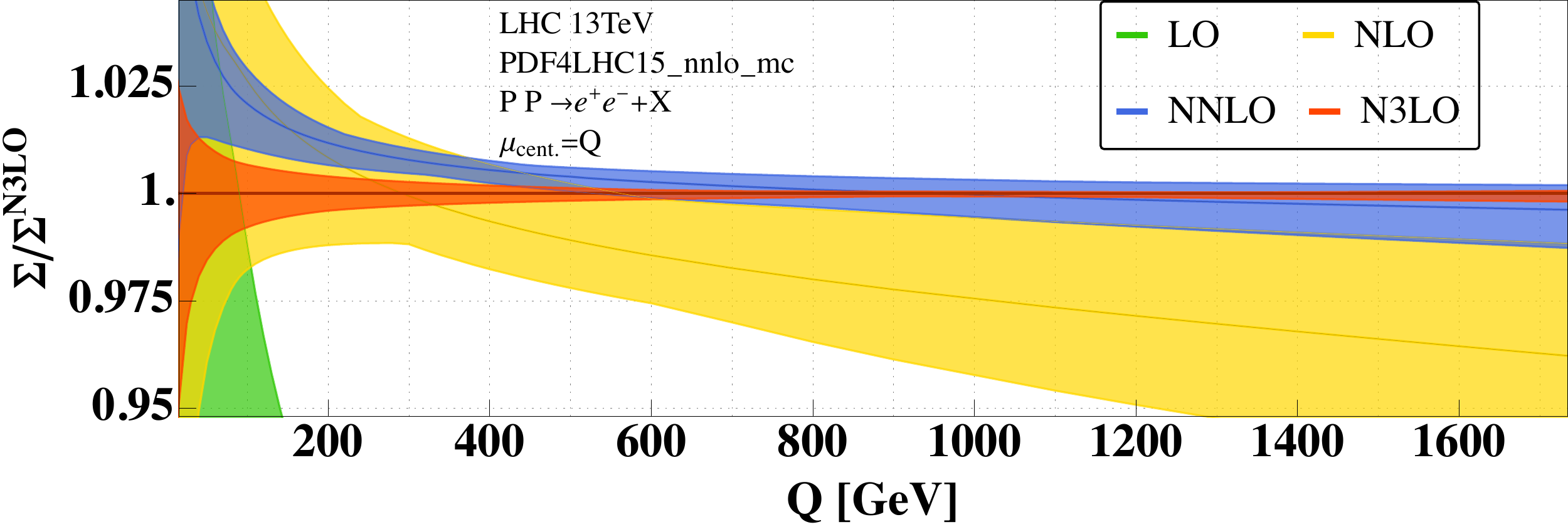}
\caption{\label{fig:totalxs}
The left panel shows the Higgs boson production cross section in gluon fusion as a function of the LHC energy through different orders in perturbation theory.
The right panel shows the invariant-mass distribution $\Sigma(Q)$ of the Drell-Yan production process at the LHC with $\sqrt{S}=13$ TeV at different orders in perturbation theory. 
}
\end{figure*}

In figure~\ref{fig:totalxs} we show the dependence of the gluon-fusion and of the neutral-current Drell-Yan (NCDY) cross sections on the perturbative scales. In the left panel, we show the Higgs cross section as a function of the center-of-mass energy $\sqrt{S}$ of a proton-proton collider at N$^k$LO, for $k=0,..,3$. The bands are obtained by varying the perturbative scales by a factor of 2 around the central scale $\mu_{\textrm{cent.}}=m_H/2$. We see that, as expected, the scale dependence $\delta(\textrm{scale})$ is reduced considerably as the perturbative order is increased, reaching a few percent at N$^3$LO. Moreover, we observe a nice convergence of the perturbative series, with the scale variation band at N$^3$LO strictly contained within the NNLO band. We stress, however, that this convergent behaviour depends on the choice of the hard scale~\cite{Anastasiou:2015ema,Anastasiou:2016cez}. 

In the right panel of figure~\ref{fig:totalxs} we show the NCDY cross section at different orders normalized to the N$^3$LO prediction as a function of the invariant mass $Q$ of the produced lepton pair. Similar to the case of Higgs production, we observe a considerable reduction of the dependence on the perturbative scales as the order is increased. At the same time, we find that the bands obtained from scale variation at NNLO and N$^3$LO do not overlap for invariant masses $60\textrm{ GeV} \lesssim Q\lesssim 400 \textrm{ GeV}$, and this conclusion is independent of the choice of the central scale. This clearly shows that care is needed when interpreting scale variation as a tool to estimate the size of the missing higher orders, especially at high orders in perturbation theory where we aim for precision predictions.

In order to investigate the relevance and the impact of N$^3$LO computations, we summarize in table~\ref{tab:summary_N3LO} the results for the inclusive production cross section for various $2\to1$ processes. All results are obtained for the LHC with $\sqrt{S}=13$ TeV, and we fold partonic cross sections with the \verb+pdf4lhc15_nnlo_mc+ set~\cite{Butterworth:2015oua}. We show results for the K-factors from NNLO to N$^3$LO, and we observe that in all cases the N$^3$LO corrections can change the value of the predictions by a few percent, up to $5\%$ depending on the invariant mass $Q$ considered. We also show the uncertainty $\delta(\text{scale})$ on the cross section from varying the perturbative scales by a factor of 2 up and down around the central scale $\mu_{\text{cent.}} = Q/2$. We see that in all cases the residual scale dependence at N$^3$LO is of the order of a few percent. Based on these results, we conclude that N$^3$LO predictions for hadron collider observables are highly desired and needed if we want to achieve percent-level precision for hadron collider observables.

\begin{table}[!h]
\begin{center}
\begin{tabular}{|c |c | c | c | c | c |}
\hline
 &$Q$ [GeV]& {K}-factor &$\delta(\textrm{scale})$ [\%]& $\delta(\text{PDF}+\alpha_S)$ & $\delta(\text{PDF-TH})$ \\
 \hline
{$gg\to ${ Higgs}} & $m_H$ &$1.04$ & ${}^{+0.21\%}_{-2.37\%}$ & $\pm 3.2 \%$ & $\pm 1.2 \%$ \\
 \hline
{$b\bar b \to ${ Higgs}} &$m_H$& 0.978 & ${}^{+3.0\%}_{-4.8\%}$ & $\pm 8.4 \%$ & $\pm 2.5 \%$ \\
 \hline
 \multirow{2}{5em}{NCDY}& 30&  0.952 & ${}^{+1.53 \%}_{-2.54 \%}$ & ${}^{+3.7\%}_{-3.8\%}$ & $\pm 2.8 \%$  \\
& 100 &  0.979 & ${}^{+0.66 \%}_{-0.79 \%}$ & ${}^{+1.8\%}_{-1.9\%}$ & $\pm 2.5 \%$  \\
\hline
 \multirow{2}{5em}{CCDY$(W^+)$}& 30& 0.953 & ${}^{+ 2.5 \%}_{-1.7\%}$ & $\pm 3.95 \%$  & $\pm 3.2 \%$ \\
& 150& 0.985  & ${}^{+0.5 \%}_{-0.5\%}$ & $\pm 1.9 \% $ & $\pm 2.1 \%$\\
\hline
 \multirow{2}{5em}{CCDY$(W^-)$}& 30& 0.950 & ${}^{+ 2.6 \%}_{-1.6\%}$ & $\pm 3.7 \%$ & $\pm 3.2 \%$ \\
& 150& 0.984  & ${}^{+0.6 \%}_{-0.5\%}$ & $\pm 2 \%$ & $\pm 2.13 \%$\\
\hline
\end{tabular}
\caption{\label{tab:summary_N3LO}Representative results for the K-factor for inclusive $2\to1$ processes at the LHC with $\sqrt{S}=13$ TeV, as well as for the main sources of uncertainty~\cite{Duhr:2019kwi,Anastasiou:2015ema,Anastasiou:2016cez,Mistlberger:2018etf,Duhr:2020seh,Duhr:2020sdp,Duhr:2021vwj}. For details, see the discussion in the main text.}
\end{center}
\end{table}

Achieving precise predictions for hadron collider processes does not only rely on our ability to perform high-order perturbative calculations, but it also requires the knowledge of the structure of the proton at the same level of precision. The latter is described by parton density functions (PDFs), which are non-perturbative quantities that need to be extracted from experimental data. Consequently, PDFs come with their own sources of uncertainty, which depend on the quality of the data and the fitting methodology used. Moreover, the value of the strong coupling constant used in the perturbative computations can only be measured from experiment or is extracted from Lattice QCD simulation. In table~\ref{tab:summary_N3LO} we show how the uncertainties on the PDFs and the strong coupling constant impact our theoretical predictions at N$^3$LO. The uncertainty $\delta(\text{PDF}+\alpha_S)$ quoted in table~\ref{tab:summary_N3LO} was computed using the recipe of~\cite{Butterworth:2015oua}. We observe that $\delta(\text{PDF}+\alpha_S)$ is always of the order of a few percent, and always significantly larger than the residual scale dependence. We also note that currently there is no PDF set available that was extracted by comparing theory and experiment at N$^3$LO accuracy, which formally introduces a mismatch into our computation. In order to assess the impact of this mismatch on our N$^3$LO predictions, we investigate the effect of evaluating the NNLO cross sections with NLO or NNLO PDFs following the recipe of~\cite{Anastasiou:2016cez}:
\begin{equation}
\delta(\text{PDF-TH}) = \frac{1}{2}\left|\frac{\Sigma^{\text{NNLO,NNLO-PDFs}}(Q) - \Sigma^{\text{NNLO,NLO-PDFs}}(Q)}{\Sigma^{\text{NNLO,NNLO-PDFs}}(Q)}\right|\,.
\end{equation}
We find that in all cases $\delta(\text{PDF-TH}) $ leads again to an uncertainty of a few percent, within the same ballpark as the dependence on the perturbative scales. We conclude that in order to achieve percent-level precision for hadron colliders, it is insufficient to just push to higher orders in perturbations theory, including the third order in the strong coupling. Instead, developments on the perturbative side must be accompanied by corresponding advances in our understanding of the PDFs.

The discussion of the previous subsection shows that, if we want to match the experimental precision from the theory side, for the future it will be high priority to push our theory predictions to include N$^3$LO corrections in the strong coupling constant (we emphasise that also electroweak and mixed-QCD electroweak corrections may be equally, and for some observables even more, important). 
The inclusive cross sections discussed in the previous section, however, are idealised observables. 
For the future, it will be important to achieve N$^3$LO precision also for differential observables with the possibility to impose arbitrary phase space cuts and also with additional jets in the final state already at LO. 
Experience shows that higher order corrections become often even more important in the presence of phase space cuts.
While the computation of inclusive N$^3$LO corrections for color-singlet production in $2\to1$ processes has recently achieved maturity, we have not yet reached the same level of understanding for differential observables involving higher-multiplicity final states. 
In the following sections we identify the most pressing challenges that need to be addressed in order to achieve N$^3$LO accuracy for a large range of key LHC processes.

%% file: Chapters/Amplitudes.tex
\section{Advances in amplitudes and new mathematical structures}
\label{sec:amplitudes}

\subsection{Multi-loop scattering amplitudes}
Multi-loop scattering amplitudes are essential building blocks for the computation of perturbative scattering cross sections. 
To be able to perform computations for $2\to2$ processes at N$^3$LO, three-loop $2\to2$ amplitudes and two-loop $2\to3$ amplitudes are required. 
While first results are becoming available~\cite{Gehrmann:2015bfy,Chicherin:2018yne,Chawdhry:2019bji,Badger:2019djh,Chawdhry:2020for,Badger:2021imn,Abreu:2018zmy,Abreu:2019odu,Abreu:2021oya,Caola:2020dfu,Chawdhry:2021mkw,Chawdhry:2021hkp,Czakon:2021mjy,Agarwal:2021grm,Agarwal:2021vdh,Caola:2021rqz,Bargiela:2021wuy,Caola:2021izf,Abreu:2021asb}, a systematic approach to the computation of the required multi-loop amplitudes is currently still missing, in part also due to a lack of understanding of the relevant special functions, see section~\ref{sec:elliptics}.
State of the art computations rely on impressive developments of efficient algorithms for the calculation of amplitudes and a continued effort to deepen our understanding of QFT.
Nevertheless, rapid growth of complexity as the loop order is increased and as problems with more massive particles are considered, renders cutting edge calculations of loop amplitudes highly non-trivial. 
To facilitate a widespread N$^3$LO and even NNLO phenomenology program a sizable amount of  scattering amplitudes that are yet unkown is required. 
Ultimately, it is necessary to find representations of amplitudes that can be evaluated efficiently and yield numerically reliable results.
What's more, the same scattering amplitudes can be useful in the computation of different processes and in different settings. Therefor it is of great interest to the community that scattering amplitudes are made publicly available and easily accessible. 

\subsection{Functions appearing in multi-loop integrals}
\label{sec:elliptics}
\begin{figure}
\begin{center}
\includegraphics[scale=0.6]{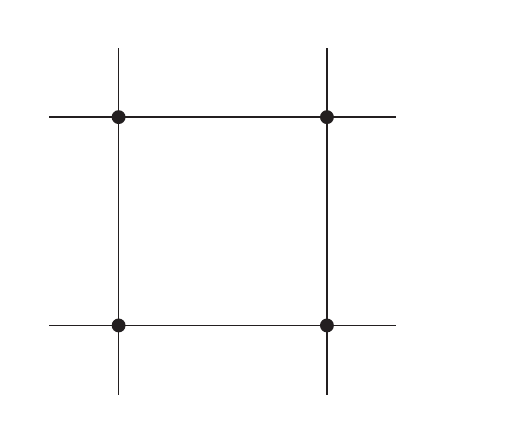}
\includegraphics[scale=0.6]{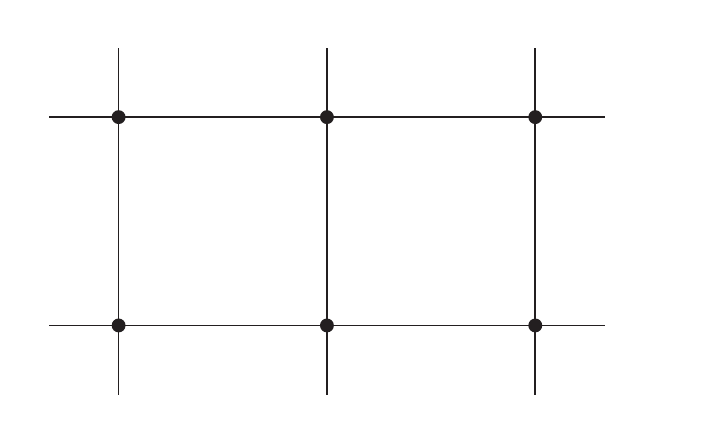}
\includegraphics[scale=0.6]{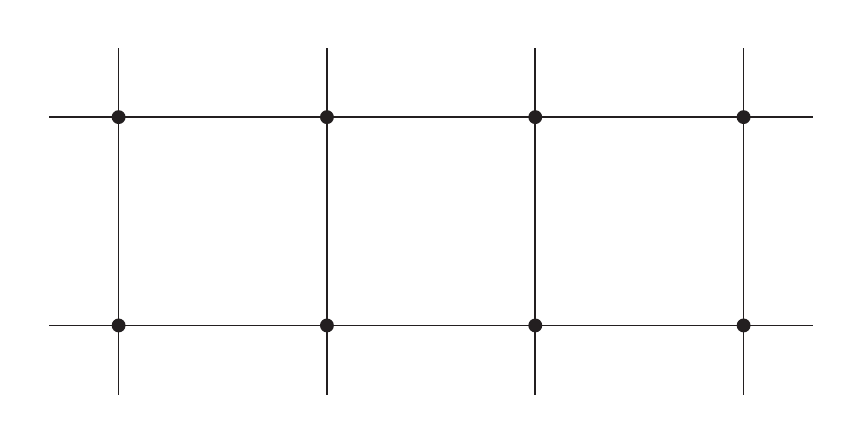}
\end{center}
\caption{
The first three members of the family of traintrack graphs.
}
\label{fig_traintrack}
\end{figure}

The Standard Model involves several heavy particles: 
the $Z$- and $W$-bosons, the Higgs boson and the top quark.
Precision studies of these particles require on the theoretical side 
quantum corrections at the two-loop order and beyond.
It is a well-known fact that starting from two-loops
Feynman integrals can no longer be expressed exclusively in terms of multiple polylogarithms.
This occurs already in two-point functions, as soon as massive particles are involved.
A prominent example is the two-loop sunrise integral with non-zero internal masses \cite{Sabry:1962,Berends:1993ee,Broadhurst:1993mw,Caffo:1998du,Laporta:2004rb,Muller-Stach:2011qkg,Adams:2013nia,Remiddi:2013joa,Bloch:2013tra,Adams:2014vja,Adams:2015gva,Adams:2015ydq,Bloch:2016izu,Adams:2017ejb,Broedel:2017siw,Bogner:2017vim,Adams:2018yfj,Bogner:2019lfa}.
\begin{figure}
\begin{center}
\includegraphics[scale=1.0]{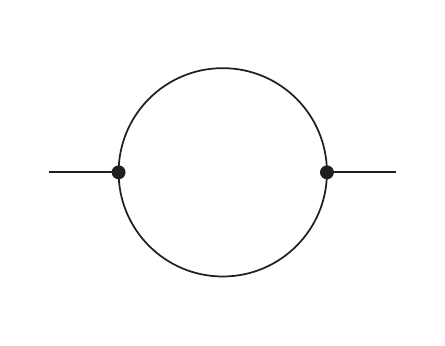}
\includegraphics[scale=1.0]{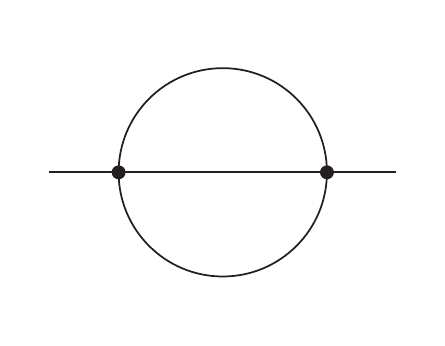}
\includegraphics[scale=1.0]{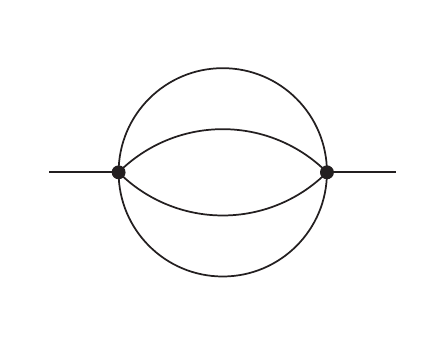}
\end{center}
\caption{
The first three members of the family of banana graphs.
}
\label{fig_banana}
\end{figure}
The sunrise integral can be viewed as a member of the family of 
banana graphs \cite{Aluffi:2008sy,Muller-Stach:2012tgj,Bloch:2014qca,Vanhove:2014wqa,Primo:2017ipr,Broedel:2019kmn,Klemm:2019dbm,Bonisch:2020qmm,Bonisch:2021yfw,Broedel:2021zij},
shown in fig.~\ref{fig_banana}.
In exploring new functions associated to Feynman integrals, we study in detail carefully chosen examples which present 
new complications in the simplest possible way.
Apart from the family of banana graphs, the family of 
traintrack graphs \cite{Bourjaily:2017bsb,Bourjaily:2018yfy,Bourjaily:2019hmc,Vergu:2020uur,Kristensson:2021ani}
(with massless propagators)
shown in fig.~\ref{fig_traintrack} is another well-studied example.
The two-loop members of these families are standard examples for elliptic Feynman integrals.

Multiple polylogarithms are associated with a complex algebraic curve of genus zero with a certain number of marked points \cite{Goncharov:2001iea,Brown:2006}.
We note that we may view any complex algebraic curve of genus $g$ as a Riemann surface of genus $g$, as one complex dimension
equals two real dimensions.
Configurations of marked points on a genus zero curve which can be obtained from another configuration by a M\"obius transformation
are considered equivalent.
The space of inequivalent configurations is called the moduli space ${\mathcal M}_{0,n}$ of $n$ marked points on a curve of genus $0$.
We may define multiple polylogarithms as iterated integrals on the moduli space ${\mathcal M}_{0,n}$, where the integrands are given
by the closed one-forms
\begin{eqnarray}
 \omega_{ij} & = & d\ln\left(z_i-z_j\right).
\end{eqnarray}
This definition may sound a little bit complicated at first sight.
It boils down to the standard definition of multiple polylogarithms in terms of iterated integrals (for $z_k \neq 0$):
\begin{eqnarray}
 G(z_1,...,z_k;y)
 & = &
 \int\limits_0^y \frac{dy_1}{y_1-z_1}
 \int\limits_0^{y_1} \frac{dy_2}{y_2-z_2} ...
 \int\limits_0^{y_{k-1}} \frac{dy_k}{y_k-z_k}.
\end{eqnarray}
The definition given above has the advantage that it paves the way towards the extension for genus one curves \cite{Weinzierl:2022eaz}.

The dimension of ${\mathcal M}_{0,n}$ is $(n-3)$, as we may use the freedom of M\"obius transformations to fix three points 
at prescribed values (usually taken as $0, 1, \infty$).
Standard coordinates on ${\mathcal M}_{0,n}$ are therefore
\begin{eqnarray}
 \left( z_1, z_2, \dots, z_{n-3} \right).
\end{eqnarray}

Starting from two-loops we also encounter transcendental functions associated to a complex algebraic curve of genus one
with a certain number of marked points.
An algebraic curve of genus one with one marked point is by definition an elliptic curve.
If we have more marked points, we speak about an elliptic curve with (additional) marked points.
We are again interested in inequivalent configurations of marked points. The space of all inequivalent configurations
is the moduli space ${\mathcal M}_{1,n}$.
The dimension of the moduli space ${\mathcal M}_{1,n}$ is $n$.
(In general, the dimension of ${\mathcal M}_{g,n}$ is $(3g+n-3)$.)
We need one coordinate to describe the shape of the elliptic curve 
(or the shape of the torus or the shape of the fundamental parallelogram). We may take the modular parameter $\tau$ for this.
We may use translation transformations to fix one marked point, say $z_n=0$.
This gives
\begin{eqnarray}
 (\tau,z_1,\dots,z_{n-1})
\end{eqnarray}
as coordinates on ${\mathcal M}_{1,n}$.
We then proceed as in the genus zero case: We consider iterated integrals on ${\mathcal M}_{1,n}$ of specific one-forms.
The one-forms we would like to consider are either of the form
\begin{eqnarray}
 \omega^{\mathrm{modular}}_{k}
 & = &
 2 \pi i \; f_k\left(\tau\right) d\tau,
\end{eqnarray}
where $f_k$ is a modular form of modular weight $k$ \cite{Adams:2017ejb}, or of the form
\begin{eqnarray}
 \omega^{\mathrm{Kronecker}}_{k}
 & = &
 \left(2\pi i\right)^{2-k}
 \left[
  g^{(k-1)}\left( z-c, \tau\right) d z + \left(k-1\right) g^{(k)}\left( z-c, \tau\right) \frac{d\tau}{2\pi i}
 \right],
 \;\;\;
\end{eqnarray}
with $c$ being a constant.
The functions $g^{(k)}(z,\tau)$ are obtained from the expansion of the Kronecker function \cite{Zagier:1991,Brown:2011,Broedel:2018qkq}
\begin{eqnarray}
 F\left(z,\alpha,\tau\right)
 & = &
 \bar{\theta}_1'\left(0,q\right) \frac{\bar{\theta}_1\left(z+\alpha, q\right)}{\bar{\theta}_1\left(z, q\right)\bar{\theta}_1\left(\alpha, q\right)}
 \; = \; 
 \sum\limits_{k=0}^\infty g^{(k)}\left(z,\tau\right) \alpha^{k-1},
 \;\;\;\;\;\;
 q \; = \; e^{2\pi i \tau}.
\end{eqnarray}
The Jacobi theta function $\bar{\theta}_1$ is defined here by
\begin{eqnarray}
 \bar{\theta}_1\left(z,q\right) 
 & \; = \; &
 -i \sum\limits_{n=-\infty}^\infty \left(-1\right)^n q^{\frac{1}{2}\left(n+\frac{1}{2}\right)^2} e^{i \pi \left(2n+1\right)z}.
\end{eqnarray}
The one-forms $\omega^{\mathrm{modular}}_{k}$ and $\omega^{\mathrm{Kronecker}}_{k}$ are closed.

Very often we solve Feynman integrals by the method of differential equations.
If we integrate on ${\mathcal M}_{1,n}$ in one of the $z$-variables we obtain elliptic multiple polylogarithms \cite{Broedel:2017kkb}:
\begin{eqnarray}
\lefteqn{
 \widetilde{\Gamma}\!\left({\begin{smallmatrix} n_1 & \dots & n_r \\ c_1 & \dots & c_r \\ \end{smallmatrix}}; z; \tau \right)
 = } & &
 \nonumber \\
 & &
 \left( 2 \pi i\right)^{n_1+\dots+n_r-r}
 I_\gamma\left( \omega^{\mathrm{Kronecker},z}_{n_1+1}\left(c_1,\tau\right), \dots, \omega^{\mathrm{Kronecker},z}_{n_r+1}\left(c_r,\tau\right); z \right),
\end{eqnarray}
with
\begin{eqnarray}
 \omega^{\mathrm{Kronecker},z}_{k}\left(c,\tau\right)
  & = & 
 \left(2\pi i\right)^{2-k} g^{(k-1)}\left( z-c, \tau\right) dz
\end{eqnarray}
being the part of $\omega^{\mathrm{Kronecker}}_{k}$ proportional to $dz$ and
$I_\gamma(\omega_1,\omega_2,\dots;z)$ denotes the iterated integral along the path $\gamma$ with endpoint $z$.

In the literature there exist various definitions of elliptic multiple polylogarithms due to the following
problem:
It is not possible that the differential one-forms $\omega$ entering 
the definition of elliptic multiple polylogarithms
have at the same time the following three properties:
(i) $\omega$ is double-periodic,
(ii) $\omega$ is meromorphic,
(iii) $\omega$ has only simple poles.
We can only require two of these three properties.
The definition of the $\widetilde{\Gamma}$-functions selects meromorphicity and simple poles.
Meromorphicity means that $\omega$ does not depend on the anti-holomorphic variables.
The differential one-forms are not double-periodic. 
This is spoiled by the quasi-periodicity of $g^{(k)}( z, \tau)$ with respect to $z \rightarrow z+\tau$.
However, this is what physics (i.e. the evaluation of Feynman integrals)
dictates us to choose.
The integrands are then either multi-valued functions on ${\mathcal M}_{1,n}$
or single-valued functions on a covering space.
Of course, in mathematics one might also consider alternative definitions, which prioritise other properties.
A definition of elliptic multiple polylogarithms, which implements properties (i) and (ii), but gives up property (iii)
can be found in \cite{Levin:2007},
a definition, which implements properties (i) and (iii), but gives up (ii) can be found in \cite{Brown:2011}.
The reader is advised to carefully check what is meant by the name ``elliptic multiple polylogarithm'', this also concerns
the definitions in \cite{Passarino:2016zcd,Remiddi:2017har}.

However, in many applications it is advantageous not to integrate in $z$ but to integrate the differential equation in $\tau$ with boundary point
$\tau = i \infty$.
We set $q=\exp(2\pi i \tau)$.
For the integration along $\tau$ we consider in $q$-space iterated integrals with integrands given by
\begin{eqnarray}
 \omega^{\mathrm{Kronecker},\tau}_{k}
 \; = \;
 \frac{\left(k-1\right)}{\left(2\pi i\right)^{k}} g^{(k)}\left(z-c, \tau\right) \frac{dq}{q}
 & \;\; \mbox{or} \;\; &
 \omega^{\mathrm{modular}}_{k}
 \; = \;
 f_{k}\left(\tau\right) \frac{dq}{q}.
\end{eqnarray}
There are two advantages:
The boundary point $\tau=i \infty$ corresponds to $q=0$. 
On the hypersurface $\tau=i \infty$ the elliptic curve degenerates to a nodal curve and we may express the
boundary values in terms of multiple polylogarithms.
The second advantage is that there will be no poles along the integration path.
There might be poles at the starting point (``trailing zeros'') or at the end point (``leading one''),
but there won't be any poles inbetween.
This is in contrast to multiple polylogarithms and elliptic multiple polylogarithms, where there might 
be poles along the integration path.
The fact that for an integration along $\tau$ there are no poles along the integration path
together with the fact that with the help of modular transformations \cite{Duhr:2019rrs,Weinzierl:2020fyx}
we may always ensure that
\begin{eqnarray}
 \left| q \right|
 & \le &
 e^{- \pi \sqrt{3}}
 \; \approx \; 0.0043
\end{eqnarray}
allows for an efficient numerical evaluation of these iterated integrals \cite{Walden:2020odh}.

Let's look at an example: The two-loop sunrise integral evaluates to iterated integrals on $\overline{\mathcal M}_{1,3}$
(the closure of $\mathcal M_{1,3}$).
In the general unequal mass case this integral depends on three dimensionless variables, which we may choose originally
as $x=-p^2/m_3^2, y_1=m_1^2/m_3^2, y_2=m_2^2/m_3^2$.
In this example the relation with algebraic geometry is simplest in the Feynman parameter representation: 
The second graph polynomial defines an elliptic curve, which intersects the domain of integration (a two-dimensional simplex)
in three points. These three points define three marked points on the elliptic curve, hence the relevant moduli space is
$\mathcal M_{1,3}$. The marked points on the curve as well as the shape of the curve vary with the kinematic variables $(x,y_1,y_2)$.
We may therefore change variables from $(x,y_1,y_2)$ to $(\tau,z_1,z_2)$ as introduced above.
In the differential equation we then find the one-forms $\omega^{\mathrm{modular}}_{k}$ and $\omega^{\mathrm{Kronecker}}_{k}$
(and only those) \cite{Bogner:2019lfa}.

Learning about new transcendental functions by studying well-chosen examples is just the first step.
In the end we are interested in physical observables.
This requires in addition efficient algorithms to express all occurring Feynman integrals in terms of our expected
function basis.
This can already be seen by looking at the two-loop electron self-energy in QED: 
In addition to the (equal mass) sunrise integral the calculation requires the kite integral \cite{Remiddi:2016gno,Adams:2016xah,Honemann:2018mrb}.

Expressing a given Feynman integral in terms of a given function basis is highly non-trivial, 
already in the case of multiple polylogarithms:
It might involve rationalisations of square roots \cite{Besier:2018jen} or matching the symbol of the Feynman integral \cite{Heller:2019gkq,Heller:2021gun}.

Thus, in addition to understand better the more complicated members of the families of banana and traintrack integrals,
efficient methods which allow us to express all Feynman integrals contributing to a particular physical observable 
at a given order in perturbation theory in terms of a known class of functions
are a direction for future research.

%% file: Chapters/Subtraction.tex
\section{IR Subtraction Schemes and their Extension to N$^3$LO}
\label{sec:subtraction}

\subsection{General Considerations}
\newcommand{\dd}{\mathrm d}

Even if all the relevant scattering amplitudes are known, obtaining
physical predictions for a given process is non trivial. This is
because QCD corrections arise from both real and virtual emissions,
which are individually infrared (IR) divergent. The divergences need
to be properly extracted and canceled to obtain a physically
meaningful results. In virtual corrections, IR singularities are
always manifest. In real contributions, however,
they only manifest themselves after integrating over QCD
radiation.  Typically one is interested in fully exclusive results,
so that such an integration cannot be performed, and a method to extract
the implicit IR singularities is required.

In general, there are at least two ways of dealing with this issue.
The first goes under the name of ``slicing'' and roughly
amounts to the following.
Since IR singularities can only be generated
in soft/collinear (SC) regions, one can separate the 
calculation of real-emission contributions in two regions: a SC
one and a hard one.
In the SC region a) QCD factorizes into a process-dependent
part that lives in the Born phase space times
universal factors and b) QCD radiation by definition does
not affect any IR-safe observable.
This allows one to integrate over extra radiation in a process- and
observable-independent way, which generates IR divergences that cancel
against the ones from virtual corrections. 
In the hard region, no IR singularities can appear
so this contribution does not pose any conceptual challenge.

To illustrate this approach,
consider Higgs production in gluon fusion at NLO. Schematically,
it can be split into virtual ($V$) and real ($R$) corrections
\begin{equation}
  \Delta\sigma^{H}_{{\rm NLO}} = \int\left[V~ \dd\Phi_{H} +
    R~ \dd\Phi_{H+1}\right],
\end{equation}
where $\dd\Phi$ is the relevant phase space. 
An effective way of separating SC and hard regions is to consider the
Higgs transverse momentum $p_{t,H}$: if it is small, QCD radiation is constrained
to be soft/collinear. One then writes
\begin{equation}
  \begin{split}
  \Delta\sigma^{H}_{{\rm NLO}} & = \int\limits_{p_{t,H}<p_{t,\rm cut}}\left[V~ \dd\Phi_{H} +
    R~ \dd\Phi_{H+1}\right] + 
    \int\limits_{p_{t,H}>p_{t,\rm cut}} R~ \dd\Phi_{H+1}.
  \end{split}
  \label{eq:higgs_slicing}
\end{equation}
If $p_{t,\rm cut}$ is small enough, one can approximate the first term on the r.h.s. of
 eq.~\eqref{eq:higgs_slicing} with its $p_{t,H}\to 0$ limit, which
is well-known from the study of small-$p_t$ resummation.  The second
term is simply the
LO Higgs transverse momentum spectrum. It is easy
to see that a similar construction generalises to arbitrary
orders: the N$^k$LO result can be written as a fully unresolved contribution
plus N$^{k-1}$LO corrections to Higgs production at finite transverse momentum. 

This examples highlights the main strengths of the slicing approach:
it allows to re-write a generic N$^k$LO calculation as the sum of a
much simpler computation which at the end only involves Born-like
configurations and a N$^{k-1}$LO calculation with one extra emission.
The main drawback is that the SC and hard contributions both develop
logarithmic sensitivity on the small $p_{t,\rm cut}$ parameter. In
general, this leads to large cancellations between different
contributions to the final result, which requires an exquisite
numerical control on the individual terms. This is non trivial for the
N$^{k-1}$LO contribution, and it comes with a heavy CPU cost. Because
of this slicing was abandoned at NLO. However, it has been 
revisited for NNLO calculations~\cite{Catani:2007vq,Boughezal:2015dva,Gaunt:2015pea}. The success of this program
is due both to advances in computing and to a progress in our
ability to perform NLO calculations very efficiently. 
Currently, slicing is the only technique that is mature enough to deal
with fully differential N$^3$LO computations, at least simple ones,
without the need for any process-dependent N$^3$LO input.  Indeed, it
has been recently applied to the calculation of N$^3$LO QCD
corrections to color-singlet production at colliders~\cite{Chen:2021vtu,Billis:2021ecs,Camarda:2021ict}. These
computations, however, are very CPU-intensive, so it is
unlikely that slicing formalisms in their current incarnation would be
able to cope with more complex N$^3$LO processes.  This issue is discussed
more in depth in sec.~\ref{sec:jettiness}, where more details on
current slicing formalisms and on progress towards their improvement
is described. 
Recently, another slicing scheme for colourful final states has been proposed in ref.~\cite{Buonocore:2022mle} using jet algorithms.

Another approach to higher-order calculations is the so-called
``subtraction'' method.  Here one does not separate between
SC and hard regions, but adds and subtracts to indivual
pieces of the computation carefully constructed counterterms that
extract and regulate the singular terms. Using the same example as
before, one schematically writes
\begin{equation}
  \Delta\sigma^{H}_{{\rm NLO}} = \int\left[V~ \dd\Phi_{H} + \mathcal S \dd\Phi_{H+1}\right]+
  \int
  \left[R-\mathcal S\right]~ \dd\Phi_{H+1}.
  \label{eq:subtr}
\end{equation}
If $\mathcal S$ reproduces $R$ in all SC configurations,
the second term of eq.~\eqref{eq:subtr} is regular in all IR
regions. Because of QCD factorisation properties, $\mathcal S$ can be
chosen in such a way that the emission of the extra parton (on top of
the Born-like configuration) is described by universal SC
functions, multiplied by a process-dependent part that lives in the
Born phase space. A similar construction can be done for the
$\dd\Phi_{H+1}$ phase space. This allows one to integrate the first
bracket of eq.~\eqref{eq:subtr} over the extra radiation, to
exposes IR singularities and cancel them against virtual corrections.

The main advantage of the subtraction approach over slicing is that 
one does not need to introduce a hard cut between SC and
hard regions, which means that no parametrically-enhanced large
cancellations between individual terms occur. The main drawback is
that it is not straightforward to re-use existing N$^{k-1}$LO
calculations for the more CPU-intensive and tedious part of the
computation.\footnote{A notable exception is the so-called
projection-to-born (P2B) approach~\cite{Cacciari:2015jma}, which however requires the
knowledge of N$^k$LO corrections fully differential in the Born
kinematics.} Also, devising a suitable $\mathcal S$ countertem and
integrating it over the unresolved phase-space (see the first term of
eq.~\eqref{eq:subtr}) are non-trivial endeavours, both conceptually
and technically.

Currently, subtractions are the standard for NLO calculations~\cite{Frixione:1995ms,Catani:1996vz,Nagy:2003qn},
and are also used as NLO input for all the NNLO slicing
results. Different subtraction formalisms have been developed for NNLO
calculations, and have been applied to several non-trivial cases. Many
frameworks can at least in principle deal with fully generic processes.
Because of the lack of large numerical cancellations, subtraction
formalisms tend to perform better than slicing, although original NNLO
formalisms were far from optimal. This led to ongoing efforts to devise
better frameworks for NNLO calculations~\cite{Gehrmann-DeRidder:2005btv,Somogyi:2005xz,Czakon:2010td,Binoth:2000ps,Caola:2017dug,Cacciari:2015jma,Sborlini:2016hat,Herzog:2018ily,Magnea:2018hab,Boughezal:2013uia}. At present, this is
still work in progress and there is no universal subtraction framework
for N$^3$LO calculation. However, 
several of the ingredients required for extending subtractions at
N$^3$LO has started to
appear~\cite{Zhu:2020ftr,Duhr:2013msa,Dixon:2019lnw,Catani:2021kcy,Catani:2019nqv} and the projection-to-born subtraction scheme~\cite{Cacciari:2015jma} could be applied to the Higgs boson production cross section in gluon fusion~\cite{Chen:2021isd}.

Both the slicing and subtraction approaches described above share the
feature of clearly separating real and virtual corrections, which are
individually IR-sensitive. In the recent past, efforts to avoid this
separation~\cite{Soper:1999xk,Catani:2008xa,Bierenbaum:2010cy} were revived, and encouraging results started to
appear~\cite{Capatti:2020xjc,Capatti:2020ytd,Aguilera-Verdugo:2019kbz}.

\subsection{N-Jettiness and $q_T$ slicing schemes}
\label{sec:jettiness}
The two main examples of slicing are the methods of $q_T$-subtraction \cite{Catani:2007vq} and $N$-jettiness subtraction \cite{Boughezal:2015dva, Gaunt:2015pea}.
In these methods we take a cross section that we want to calculate, let's call it $\sigma(X)$, and we rewrite it as the integral of a differential distribution with respect to an observable $\tau$ such that we can split its calculation in two regions as 
\beq\label{eq:slicingstep1}
	\sigma(X)= \int \df \tau \frac{\df \sigma(X)}{\df \tau}= \int_0^\taucut \df \tau \frac{\df \sigma(X)}{\df \tau} + \int_\taucut \df \tau \frac{\df \sigma(X)}{\df \tau}\,.
\eeq 
Here, $\sigma(X)$ can be a differential cross section and we identify with $X$ the set of measurements that we want to consider, which can include experimental cuts and constraints on the kinematics of particles, such as the rapidity of the Higgs or an electroweak boson.
If we choose the observable $\tau$ appropriately, the region \emph{above the cut} $\tau \geq\taucut>0$, i.e. the second term in \eq{slicingstep1}, is populated only by events where there is at least one additional particle with respect to the Born configuration we are considering.
This implies that, if $\sigma(X)$ is an $n$-loop cross section for an $m$-jet process, the calculation above the cut will involve the cross section for $m+1$-jet process at $n-1$ loops.
Therefore, the true $n$-loop singularities arise only in the region \emph{below the cut}, $\tau <\taucut$, in particular for $\tau \to 0$.
We can then expand the differential cross section around the singular limit $\tau \to 0$ as 
\beq
	\frac{\df \sigma(X)}{\df \tau} = \frac{\df \sigma^{\sing}(X)}{\df \tau} + \sum_{i>0} \frac{\df \sigma^{i}(X)}{\df \tau} \,,
\eeq
where the first term contains the singular behavior scaling as $\frac{1}{\tau}$, while the rest of the terms $\frac{\df \sigma^{i}(X)}{\df \tau}$ are suppressed by powers of $\tau^i$ with respect to the leading power term and therefore are integrable.
With that, we may recast \eq{slicingstep1} as
\beq\label{eq:slicingstep2}
	\sigma(X)= \int_0^\taucut \df \tau \frac{\df \sigma^\sing(X)}{\df \tau} + \int_\taucut \df \tau \frac{\df \sigma(X)}{\df \tau} + \Delta\sigma(X,\taucut)\,,
\eeq 
where 
\beq
	\Delta\sigma(X,\taucut) \equiv \sum_{i>0} \int_0^\taucut \df \tau \frac{\df \sigma^i(X)}{\df \tau}\,.
\eeq
Since the terms in $\Delta\sigma(X,\taucut)$ are integrable, we have that $\Delta\sigma(X,\taucut) \to 0$ as we take $\taucut \to 0$.
Knowing just the leading power term $\frac{\df \sigma^\sing(X)}{\df \tau}$ below the cut and the $n-1$ loop cross section above the cut, we have a way of obtaining $\sigma(X)$ with an error $\Delta\sigma(X,\taucut)$ that vanishes as $\taucut\to 0$.
Note, that this method can be refined to obtain a subtraction that is local in $\tau$ where the $\cO(1/\tau)$ divergences cancel locally at the integrand level, see \cite{Gaunt:2015pea,Billis:2019vxg} for details. This is implemented, for example, in the parton shower program {\tt Geneva}~\cite{Alioli:2012fc, Alioli:2013hqa, Alioli:2015toa} for the $0-$jettiness variable.

The $N$-jettiness and $q_T$ subtraction methods have been successfully applied to a large variety of differential calculations at NNLO for the LHC \cite{Grazzini:2008tf, Catani:2009sm, Grazzini:2017mhc, Boughezal:2015dva, Boughezal:2015ded, Boughezal:2016yfp,Boughezal:2016isb, Boughezal:2016dtm, Campbell:2016jau, Boughezal:2016wmq, Boughezal:2016wmq, Heinrich:2017bvg, Campbell:2019gmd} and recently at N$^3$LO \cite{Billis:2021ecs,Chen:2021vtu}, hence constituting one of the backbones of the modern toolbox for generating high precision predictions for collider observables. 
However, both formally and practically, one cannot take $\taucut$ to be arbitrarily small, as the cross section above the cut is divergent for $\taucut \to 0$ and so one always have a residual error due to $\Delta\sigma(X,\taucut) \neq 0$. 
Substantial progress have been made in obtaining analytic control of $\Delta\sigma(X,\taucut)$ by studying perturbative power corrections\cite{Moult:2016fqy, Moult:2017jsg, Ebert:2018lzn,Boughezal:2016zws, Boughezal:2018mvf, Ebert:2018gsn, Bhattacharya:2018vph, Boughezal:2019ggi,Ebert:2019zkb,Ebert:2020dfc,Vita:2020ckn, Cieri:2019tfv,Buonocore:2019puv}.
Note in particular that there can be a strong dependence on the precise definition of the observable, see for example the discussion on the leptonic definition of $\Tau_0$ in ~\cite{Moult:2016fqy, Moult:2017jsg, Ebert:2018lzn, Boughezal:2019ggi}. 

Crucially, the leading power term $\frac{\df \sigma^\sing(X)}{\df \tau}$ is very well understood and it is described by factorization theorems using effective field theory such as Soft and Collinear Effective Theory (SCET) \cite{Bauer:2000ew, Bauer:2000yr, Bauer:2001ct, Bauer:2001yt}.

The leading power factorization theorem for $q_T$ reads \cite{Collins:1981uw,Collins:1984kg,Catani:2000vq,deFlorian:2001zd,Becher:2010tm, Chiu:2012ir,Echevarria:2012js}
\begin{align} \label{eq:TMD_factorization}
 \frac{\df\sigma^\sing}{\df Q^2 \df Y \df^2 \qt} &
 = \sum_{a,b} H_{ab}(Q^2,\mu) \int\!\frac{\df^2\bt}{(2\pi)^2} e^{\img\,\qt \cdot \bt}
   \, B_a\Bigl(x_1^B, b_T, \mu, \frac{\nu}{\omega_a}\Bigr)
   \, B_b\Bigl(x_2^B, b_T, \mu, \frac{\nu}{\omega_b}\Bigr)
   \, S(b_T, \mu, \nu)
\nn\\&
 = \sigma_0 \sum_{a,b} H_{ab}(Q^2,\mu) \int\!\frac{\df^2\bt}{(2\pi)^2} e^{\img\,\qt \cdot \bt}
   \,\tilde f_a(x_1^B, b_T, \mu, \zeta_a) \, \tilde f_b(x_2^B, b_T, \mu, \zeta_b)
\,,\end{align}
where $\bt$ is Fourier-conjugate to $\qt$ which is the commonly used notation in the literature, as the functions factorize as simple products in impact parameter space (see \refcite{Ebert:2016gcn} for a formulation in momentum space).
Analogously, the singular limit of $0$-Jettiness, which is relevant for inclusive and differential color singlet production at the LHC, takes the form
\begin{align} \label{eq:Tau0_factorization}
 \frac{\df\sigma^\sing}{\df Q^2 \df Y \df \Tau_0} &
 = \sum_{a,b} H_{ab}(Q^2, \mu) \int \! \df t_a \, \df t_b \,
   B_a(t_a, x_a, \mu) \, B_b(t_b, x_b, \mu) \,
   S_c\Bigl(\Tau_0 - \frac{t_a}{Q_a} - \frac{t_b}{Q_b}, \mu\Bigr)
\,.\end{align}
For the application of $q_T$ or $N$-jettiness subtraction at N$^n$LO, \emph{all} ingredients appearing in the leading power factorization theorem must be calculated at N$^n$LO, so let us give some details on the objects appearing in  \eqs{TMD_factorization}{Tau0_factorization}.
We take $Q^2 = x_1^B x_2^B E^2_{\rm cm}$ and $Y = \frac{1}{2}\log\left(\frac{x_1^B}{x_2^B}\right)$ to be the color singlet invariant mass and rapidity, respectively.
The hard function $H_{a,b}$ encodes the virtual corrections to the born process $a,b \to h$, with $h$ being the color singlet particle. It is related to the square of the IR finite part of the quark/gluon form factor which is known up to N$^3$LO for more than a decade \cite{Kramer:1986sg, Matsuura:1987wt, Matsuura:1988sm, Gehrmann:2005pd, Moch:2005tm, Moch:2005id, Baikov:2009bg, Lee:2010cga, Gehrmann:2010ue} and has now been calculated at N$^4$LO in \refcite{Lee:2022nhh}. For the explicit expressions for $H$ itself up to 3 loops. see \refcite{Gehrmann:2010ue}.
In \eq{TMD_factorization}, the soft function $S(b_T, \mu, \nu)$ encodes the information about the QCD soft radiation dynamics in the presence the transverse momentum measurement constraint and can be formulated as a shifted Wilson line matrix element in position space, see \refcite{Li:2016axz} for details, and it is known at N$^3$LO \cite{Li:2016ctv}.
$B_a\Bigl(x_1^B, b_T, \mu, \frac{\nu}{\omega_a}\Bigr)$ is the $q_T$ beam function for the parton $a$ and it describes the probability of finding such parton in the proton with longitudinal momentum fraction $x$ and impact parameter $\bt$.
The beam functions allow an operator product expansion for $q_T \gg \lqcd $ onto standard longitudinal pdfs in terms of perturbatively calculable matching kernels
\begin{align} \label{eq:beam_OPE_schematic}
 B_i(z, \bt,\mu,\nu) = \sum_j \int_z^1\!\frac{\df z'}{z'} \, \cI_{ij}(z',\bt,\mu,\nu) f_j\Bigl(\frac{z}{z'},\mu \Bigr) \times \bigl[1 + \cO(b_t \lqcd)\bigr]
\,.\end{align}
The kernels $\cI_{ij}(z',\bt,\mu,\nu)$ have been recently calculated up to N$^3$LO in \refcite{Luo:2019szz,Ebert:2020yqt} both for the quark and the gluon case. 
This provided the last missing ingredients to apply $q_T$ subtraction at N$^3$LO and have been implemented for fiducial Higgs production in \refcite{Billis:2021ecs} and differential rapidity distribution for Drelly-Yan in \refcite{Chen:2021vtu}.
Note that due to the well known presence of rapidity divergences \cite{Collins:1981uk,Ji:2004wu,Beneke:2003pa, Chiu:2007yn, Becher:2011dz,Chiu:2011qc, Chiu:2012ir,Chiu:2009yx, GarciaEchevarria:2011rb,Li:2016axz,Ebert:2018gsn} the beam and soft functions separately depend on a rapidity scale $\nu$, but can be combined, as we have done in the second line of \eq{TMD_factorization}, into Transverse Momentum Dependent Parton Distribution Functions (TMDPDFs), 
\beq\label{eq:TMDPDFdef}
 f_i(x, b_T, \mu, Q) = B_i\Bigl(x, b_T, \mu, \frac{\nu}{Q}\Bigr) \sqrt{S(b_T, \mu, \nu)}
\,,\eeq
which are rapidity renormalization scheme independent objects.
\subsection{The N-jettiness soft function at N$^3$LO}
The $q_T$ subtraction method has been successfully applied to color singlet production processes at N$^3$LO for the LHC.
To also describe the production of colored final states, a different method is required to isolate the initial and final state singularities simultaneously. 
Such a method is provided by the $N$-jettiness slicing method.
When final state colored objects are computed, such as in jet production or in the case of the Higgs $p_T$ spectrum, the singular limit of the cross section in the regime where $\Tau_N < \tau_{\rm cut} $ takes the factorized form 
\begin{align} \label{eq:TauN_factorization}
 d\sigma(\Tau_N < \tau_{\rm cut})&
 =  H  \int B_a \otimes B_b \otimes \sum_{n}^N J_n \otimes S_{N} \,,
\end{align}
where in addition to the beam function $B_a$ in the $0$-jettiness case, $J_n$ is the inclusive jet function to take care of the final state collinear singularities. $S_N$ is the soft function which divides the phase space into $N$ different partitions, each with one and only one collinear singularity. The partition of the phase space follows the definition of the $N$-jettiness 
\begin{align} \label{eq:tauN_def}
\Tau_N = 
\sum_k \min_i \left\{ 
2 \frac{p_i \cdot q_k}{Q_i}
\right\} \,. 
\end{align} 
 The $p_i$ are light-like vectors for each of the
initial beams and final-state jets in the problem, while the
$q_k$ denote the four-momenta of any final-state QCD radiation.
The $Q_i$ are dimensionful variables that characterize the
hardness of the beam-jets and final-state jets. 

To apply eq.~(\ref{eq:TauN_factorization}) to N$^3$LO calculations, all ingredients therein have to be calculated to $3$-loops. 
The quark jet function was first calculated to $3$-loops in~\cite{Bruser:2018rad} and was confirmed in~\cite{Banerjee:2018ozf} by a different method that uses
the three-loop coefficient functions~\cite{Vermaseren:2005qc,Soar:2009yh} for deep-inelastic scattering via the exchange of a virtual
photon that couples to quarks.
When couple the photon to a scalar instead, it has also been used to obtain
the gluon jet function~\cite{Banerjee:2018ozf}.  
The beam function $B_a$ is also fully known to N$^3$LO for both the quark and gluon case~\cite{Ebert:2020unb}, by means of the collinear expansion strategy of the differential Higgs boson and Drell-Yan production cross sections~\cite{Ebert:2020lxs}.

Compared with the jet and beam functions, the  soft function is currently known only at NNLO, partly due to the existence of the step $\theta$-functions to partition the phase space in the soft function. At NNLO, the soft function was obtained numerically by means of the phase space sector decomposition method~\cite{Boughezal:2015eha,Jin:2019dho}, thanks to the perfect knowledge of the real-virtual (RV) soft current at ${\cal O}(\alpha_s^2)$ with no $\epsilon$-expansion~\cite{Catani:2000pi}. The strategy, however, may likely not work at N$^3$LO. The calculations of the double-real-virtual and double-virtual-real correction is more complicated and the $\epsilon$-expansion seems inevitable in order to carry out the loop integrals. Though challenging, recent developments for the computation of phase-space integrals with step functions~\cite{Chen:2019mqc,Chen:2019fzm,Chen:2020wsh,Baranowski:2021gxe} bring new opportunities to obtain the $\alpha_s^3$ order soft function analytically. 

%The phase space integration involves the experimental cuts, and are usually represented by the theta functions. 
Due to the presence of theta functions, some modern techniques of the loop calculations, such as the integration-by-parts (IBP) method \cite{Tkachov:1981wb,Chetyrkin:1981qh} and the differential-equation method \cite{Kotikov:1990kg,Remiddi:1997ny}, cannot be directly applied. 
Theta functions appear in many instances, for example as experimental constraints.
One example is the $N$-jettiness soft functions, for which at the moment only numerical results are known at $2$-loops. 
The problem of step functions in phase space integrals can be solved by using the method developed in refs.  \cite{Chen:2019mqc,Chen:2019fzm,Chen:2020wsh}.
The idea is that a theta function has an integral representation similar to the alpha parametrization of a propagator. 
Thus integrals with theta functions can be reduced and calculated by constructing IBP identities and differential equations in the parametric representation.

Specifically,the step $\theta$-function and the $\delta$-function belong to the integral class with definition
\begin{equation}
w_\lambda(u)\equiv e^{-\frac{\lambda+1}{2}i\pi}\int_{-\infty}^{\infty}\mathrm{d}x\frac{1}{x^{\lambda+1}}e^{ixu}.
\end{equation}
 Obviously, we have $w_0(u)=2\pi\theta(u)$ and $w_{-1}(u)=2\pi\delta(u)$. Thanks to the similarity with the alpha parametrization of a propagator, the function $w_\lambda$ can be regarded as a propagator with an index $\lambda$. Thus integrals with theta functions can directly be parametrized. The resulting parametric integrals $I(\lambda_1,\lambda_2,\dots,\lambda_n)$ satisfy the identities

\begin{equation}\label{eq:ParIBP}
\left(D_0\frac{\partial\mathcal{F}(\hat{x})}{\partial \hat{x}_i}-\hat{z}_i\right)I(\lambda_1,\lambda_2,\dots,\lambda_n)=0,\quad i=1,~2,\dots,n+1,
\end{equation}

\noindent where $D_0$ is an operator that increases the spacetime dimensions by $2$, and $\hat{x}$ and $\hat{z}$ are operators increasing or decreasing the indices.\footnote{For a regular propagator, $\hat{x}$ increases the index and $\hat{z}$ decreases the index. For a ``propagator'' $w_\lambda$, $\hat{x}$ decreases the index and $\hat{z}$ increases the index.} $\mathcal{F}$ is related to the well-known Symanzik polynomials $U$ and $F$ through $\mathcal{F}(x)\equiv F(x)+U(x)x_{n+1}$.

Based on eq.~(\ref{eq:ParIBP}), two methods to reduce the parametric integrals were presented in ref. \cite{Chen:2019fzm}. Meanwhile, differential equations can also be constructed straightforwardly in the parametric representation, and therefore the techniques for loop calculations can be systematically applied to the phase space integrals with theta functions. The computational framework developed will make the analytic calculations of many observables at N${}^3$LO possible, such as the thrust or $N$-jettiness soft functions.

As an example, we consider the calculation of the following integral:

\begin{equation*}
I=\frac{(2\pi)^5}{\pi^{3d/2}}\int\mathrm{d}^dl_1\mathrm{d}^dl_2\mathrm{d}^dl_3\frac{\delta(l_1^2)\delta(l_2^2)\delta(l_1^--l_1^+)\delta(l_1^++l_2^+-1)\theta(l_2^--l_2^+)}{l_3^2(l_3+l_1)^2(l_3-l_2)^2(-l_2^+)(l_1^++l_3^+)(l_2^--l_3^-)}.
\end{equation*}

\noindent This integral is relevant for the calculation of the RRV contribution to the thrust soft function at N$^3$LO. Here we use the lightcone coordinates. That is, $l_i^+\equiv l_i\cdot n_1$, $l_i^-\equiv l_i\cdot n_2$, with $n_1^2=n_2^2=0$, and $n_1\cdot n_2=2$. To construct differential equations, we insert an auxiliary delta function $\delta(l_1^-+l_2^--y)$. Following the approach that we have presented, the resulting integral can then be calculated by the standard differential-equation method. The result reads

\begin{equation*}
\begin{split}
I=&ie^{i\pi\epsilon}[2\pi\Gamma(1+\epsilon)]^3\left[-\frac{11}{12 \epsilon^4}-\frac{7}{3 \epsilon^3}+\frac{1}{\epsilon^2}\left(\frac{29}{3}+\frac{13 \pi ^2}{36}\right)+\frac{1}{\epsilon }\left(\frac{59 \zeta_3}{2}-\frac{121}{3}-\frac{13 \pi^2}{9}\right)\right.\\
&\left.+\frac{509}{3}+\frac{41 \pi ^2}{9}+\frac{1091 \pi ^4}{360}-4 \pi ^2\log (2)\right].
\end{split}
\end{equation*}

%% file: Chapters/Challenges.tex
%\newpage
\section{Bottlenecks and open questions}
\label{sec:challenges}

Achieving the goal of percent level physics in hadron collisions will require a concerted effort and support from particle physics phenomenology community.
In previous sections, we discussed the status and significant developments towards perturbative corrections at N$^3$LO.
Below we identify some of the mayor challenges that have to be overcome in the future.

\subsection{ Accessibility and User Friendliness} 
Creating frameworks that make N$^3$LO (and NNLO) predictions easily accessible to a large community should be a priority. 
An enormous degree of automatisation and easy access for NLO computations lead to the current back-bone of modern high energy LHC phenomenology. 
Repeating a similar success story at higher orders is important to achieve our scientific goals.
Predictions at N$^3$LO will reach a much larger audience and unfold their impact fully if they can be accessed in terms of public software and such infrastructure is support by the community. 
For example, individual publications by members of the theoretical community may predict a particular observable as measured by a particular experiment. This or another experiment may in the future decide to modify the definition of the observable slightly by varying simple input parameters. 
If the prediction of an observable relies on public software and the experimental analysis team can easily adjust the software, then the original theoretical work will be useful for the new analysis. 
Other examples are the demand for easy access in the extraction of PDFs, the development of new observables or the comparisons and validation of new computations.
The benefit of disseminating theoretical results in an optimal way cannot be overstated but also not fully explored in the scope of this article.

\subsection { Corrections beyond QCD} 
The computation of QCD corrections to scattering cross sections is one particular perturbative input towards precision predictions. 
Corrections due to the exchange of electroweak particles are typically of similar size as NNLO corrections.
This is most easily illustrated by comparing the electroweak and strong coupling constant, $\alpha_S^2 \sim \alpha_{\text{EWK}}$. 
This implies $\alpha_S^3 \sim \alpha_{\text{EWK}}\alpha_S$, so that for high precision observables even mixed QCD-electroweak corrections will become important. 
Equally important may be effects due to non-vanishing quark masses.
As a first and often good approximation light quark masses are neglected in perturbative computations.
However, when experimental precision becomes high enough these effects need to be accounted for.
The technological challenges for the computation of electroweak and mass corrections are very similar to those of performing N$^3$LO computations and by large the statements of previous sections apply.
It should be noted that aiming for higher precision in our predictions does not only complicate computations due to the perturbative complexity of gauge theory at increasing perturbative order but also due to the fact that many more ingredients like electroweak and mass corrections need to be taken into account.

\subsection { Factorisation Violation at N$^3$LO and beyond}
Apart from developing formalisms for N$^3$LO calculations, it is also
important to investigate whether the standard perturbative approach is
sensible at all. Studies in this direction are in their infancy, but
will be crucial to the precision program. Currently, it is already
known that a naive perturbative approach would not work for top
production at N$^3$LO~\cite{Beneke:2016jpx,Melnikov:2014lwa}, and there are indications that the
standard framework of collinear factorisation may become less
straightforward at higher orders~\cite{Catani:2011st,Forshaw:2006fk,Dixon:2019lnw}. Another important issue is
whether uncontrolled non-perturbative effects may be actually larger
than higher-order perturbative corrections, making the latter
pointless. In the recent past, there was some indication that this may
not be the case for several key collider observables~\cite{Caola:2021kzt} but more
work is needed to establish a clear picture.

\subsection { DGLAP-evolution at four loops} We have already emphasised that the development of tools for perturbative computations at N$^3$LO goes hand in hand with the need to improve our understanding of the PDFs describing the internal structure of the proton. PDFs are non-perturbative and need to be extracted from experimental data. The evolution of the PDFs with the factorization scale is purely perturbative and is governed by the celebrated Dokshitzer-Gribov-Lipatov-Altarelli-Parisi (DGLAP) equation~\cite{Altarelli:1977zs,Dokshitzer:1977sg,Gribov:1972ri}. The DGLAP anomalous dimensions are currently known to three loop order~\cite{Ablinger:2014nga,Ablinger:2017tan,Moch:2004pa,Vogt:2004mw}, which is sufficient for NNLO computations. In order to achieve N$^3$LO accuracy for PDFs, the four-loop corrections to the anomalous dimensions will be required and an effort of their computation is on-going~\cite{Moch:2017uml,Moch:2018wjh,Moch:2021qrk}. Their full computation is technically very challenging, and will benefit from the developments discussed in section~\ref{sec:amplitudes}.

\subsection { N$^3$LO PDFs} 
As more and more computations for cross sections at N$^3$LO become available, it will be possible to use these computations to extract parton distribution functions at this order from a global data set. 
The importance of precise N$^3$LO PDFs was already discussed in section~\ref{sec:status}.
In particular, moving towards N$^3$LO PDFs will be associated with additional conceptual challenges, like the treatment of theoretical uncertainties or the simultaneous extraction of quark masses or the strong coupling constant. 
Furthermore, a rise in complexity can be expected to go along side an increase in the demand for computing facilities for PDF extraction.
We would like to stress that fits of PDFs are highly non-trivial and require significant community support. 
Developing and maintaining the infrastructure for the PDF extraction should be a priority.

\subsection { Parton Showers} 
Parton shower Monte Carlo generators are the workhorses of the LHC phenomenology community.
They allow for the simulation of realistic scattering events in LHC detectors. 
A consistent combination of parton showers with fixed order perturbative computations at high perturbative orders is a challenging task but should be envisioned for the future of the precision program at the LHC.
Currently, NLO accurate parton showers are wide spread and successes at NNLO are available (see for example refs. ~\cite{Monni:2019whf,Alioli:2021qbf,Mazzitelli:2021mmm,Lombardi:2021rvg,Alioli:2021egp}).
Recently, extensions to combine parton showers and N$^3$LO computations have been proposed~\cite{Prestel:2021vww,Bertone:2022hig}.
The fusion of N$^3$LO calculation and parton shower Monte Carlo event generators should be further explored and would yield a particularly efficient way of making high order precision computations useful to the larger community.

\subsection { Resummation}
Fixed order computations fail to describe scattering processes adequately in soft and collinear regions of phase space.
To adequately describe such regions, the perturbative series can be re-ordered and observables can be described to all perturbative orders but limited logarithmic accuracy. 
Such resummation techniques are widely available and an extremely successful tool in the standard tool box of high energy physics. 
Their futher development in conjunction with high order computations is important.
In particular, the existence of higher order computations often leads to the extraction universal ingredients necessary for ever more precise resummed predictions.
The combination of resummation and fixed order predictions then allows for a precise description of observables in and away from soft and collinear sensitive regions.

\subsection { Uncertainties} 
The truncation of the perturbative expansion naturally leads to an incomplete description of fundamental scattering processes and an uncertainty has to be associated with the prediction of an observable up to a given order. 
Largely, our field currently uses variations of unphysical scales to estimate such an uncertainty. 
This method is  heuristic and lacks a proper statistical interpretation. 
As theory uncertainties become dominant, their quantitative assessment becomes more important. 
Furthermore, explicit frameworks to derive statistically sound statements about theoretical uncertainties and correlations of such uncertainties among many observables will play an increasingly important role.
Steps to a more refined formulation of such uncertainties have been taken~\cite{Bonvini:2020xeo,Bagnaschi:2014wea,Duhr:2021mfd} but more research in this direction is desirable.

% \subsection { Beyond Leading Power Factorisation:} Exploring the limitations of leading power perturbative descriptions of hadron collision cross sections.

%% file: Chapters/Conclusions.tex
\section{Conclusions}
\label{sec:conclusions}

In this article we describe one particular requirement towards percent precision phenomenology for LHC processes: the ability to perform computations at N$^3$LO in perturbative QCD.
We study the status of current, inclusive N$^3$LO predictions for idealised observables and find the following:
\begin{enumerate}
\item N$^3$LO corrections for typical LHC processes are at the level of several percent and thus important to reach the LHC target precision.
\item The inclusion of N$^3$LO corrections overall significantly improves the description of observables.
\item Residual uncertainties due to the truncation of the perturbative expansion are at the percent level.
\item Residual perturbative uncertainties are of comparable size to other typical uncertainties associated with perturbative predictions.
\end{enumerate}

Next, we explore key ingredients towards enabling future N$^3$LO computations for a wide range of realistic predictions of observables. 
We highlight the importance of the computation of multi-loop scattering amplitudes as a key ingredient to N$^3$LO cross sections. 
We emphasise the importance of the development of techniques and tools for the computation of such amplitudes. 
In particular, we highlight the impact of the development of our mathematical understanding of the functions comprising scattering amplitudes. 
So-called elliptic multiple polylogarithms are representatives of functions that are part of active mathematical and physics research and play a key role in as building blocks for scattering amplitudes.

Another key aspect of future N$^3$LO predictions is the development of efficient strategies and algorithms for the integration of final state degrees of freedom. 
In particular, we discuss the current status of slicing and subtraction techniques and explore their applicability to N$^3$LO calculations. 
We emphasize the importance of future developments of such phase space integration techniques to achieve a broad range of fully realistic predictions at N$^3$LO.

We highlight that achieving percent level precision for processes of high phenomenological relevance will rely on much more than QCD corrections at N$^3$LO. 
We discuss briefly some of the theoretical developments we envision to go alongside the further development of predictions at N$^3$LO in QCD.

The precision phenomenology program of LHC has the enormous potential to answer some of the most pressing question of contemporary physics. 
The success of this program relies on our capability to interpret the observed data. 
Being able to perform accurate and precise predictions using perturbative quantum field theory is a key element in this process. 
In particuar, predictions at N$^3$LO in QCD will play the role of a future precision standard for important LHC observables.
Developing the understanding and technology to realise widespread N$^3$LO phenomenology represents an exciting and multi-facetted field of active research.